\documentclass[prl,twocolumn,showpacs,floatfix,amsfonts]{revtex4}
\usepackage{graphicx,graphics,color,epsfig}
\usepackage{bm}
\usepackage{amsmath}
\usepackage{amssymb}

\begin{document}
\preprint{}

\title{Spatial Distribution of LDOS in Cuprate Superconductors with
Magnetic-Field-Induced Stripe Modulations}
\author{Hong-Yi Chen and C.S. Ting}
\affiliation{Texas Center for Superconductivity and Advanced Material,
and Department of Physics, University of Houston, Houston, TX 77204}

\begin{abstract}
A phenomenological model defined in a two dimensional lattice is employed 
to investigate the $d$-wave superconductivity and the competing
antiferromagnetic order in cuprate superconductors. Near the optimally
doped regime, we show that it is possible to induce the spin density wave 
(SDW) and the accompanying charge density wave (CDW) orders with stripe 
modulations by applying a magnetic field.
The periods of the magnetic field induced SDW and CDW are $8a$ and $4a$,
respectively. The spatial profiles of the local density of
states (LDOS) at various bias energies have also been numerically studied.
Near and beyond the energies of the vortex core states, we found that the
LDOS may display the CDW stripe-like modulation throughout the whole 
magnetic unit cell. For energies closer to the zero bias, the stripes 
appear to be rather localized to the vortex. The intensity of the 
integrated spectrum of the LDOS shows that the strength of the stripe 
modulation is decaying away from the vortex core. This feature is in good 
agreement with STM experiments. The case for the magnetic-field induced 
SDW/CDW with 4-fold symmetry has also been studied.
\end{abstract}


\maketitle

In hole-doped cuprate superconductors, the interplay between the $d$-wave
superconductivity (dSC) and antiferromagnetic (AF) order has been studied
extensively in the literatures. Inelastic neutron scattering experiments
showed the presence of incommensurate spin structures in
La$_{2-x}$Sr$_x$CuO$_4$ (LSCO) \cite{lake291} with a spatial periodicity
$8a$ in the presence of a magnetic field. In addition, NMR imaging
experiment on YBa$_2$Cu$_3$O$_{7-x}$ (YBCO) observed a strong AF
fluctuation outside the vortex \cite{mitrovic413} indicating the possible
existence of spin-density-wave (SDW) gap inside the vortex. A recent STM
experiment by Hoffman $et\;al.$ \cite{hoffman295} seems to confirm the
coexistence of the static charge modulation and the superconductivity on
Bi$_2$Sr$_2$CaCu$_2$O$_8$ (BSCCO) under a magnetic field. The authors 
reported that a four unit cell checkerboard pattern is localized in a 
small region around the vortex, and its intensity is exponentially 
decaying away from the vortex core. All these experiments indicate that 
the AF fluctuations could be pinned by the vortex cores and they may form 
a static SDW/charge-density-wave (CDW) like modulations in certain samples 
of cuprate superconductors. Theoretically, a number of works proposed that 
the observed checkerboard patterns could be explained by the SDW order 
with the two-dimensional (2D) \cite{handong89,polkovnikov65,zhu89} or 
stripe \cite{chen66,podolsky67,ichioka70} modulations.
Both the 2D- and stripe-SDW 
orders induced by the magnetic field are well known to have the 
accompanying CDW modulations. In particularly, the checkerboard pattern 
has been attributed to the superposition of stripe modulations of the CDW
\cite{chen66,kivelson0210} oriented along $x$- and $y$- directions. In all
these studies, the STM spectra obtained from the experiment have been
directly interpreted in terms of the CDW order induced by the magnetic
field. This is because that the conventional wisdom leads us to believe
that the symmetry of the CDW order should be reflected in the STM spectra.
On the other hand, the CDW order represents the charge density 
configuration in the ground state while the STM spectrum or the spatial
distribution of the local density of states (LDOS) is determined by the
behavior of the low-energy excitations in the system. It is therefore
extremely interesting to know the difference between the CDW order and the
spatial profile of the LDOS. This issue has not yet been addressed in the
literature, and thus is the main purpose of the present paper.

The phonemenological $t-t'-U-V$ model defined in a 2D lattice and the
Bogoliubov-de Gennes' equations will be used to numerically study the
interplay between the $d$-wave superconductivity (dSC) and the competing 
AF order for samples close to the optimal doping. First the phase diagrams 
for the coexistence between these two orders as a function of $U/V$ in 
both zero and finite magnetic field are going to be examined. Two 
different values of $U/V$ will be chosen such that their magnitudes yield 
only the dSC when the magnetic field is zero. In the presence of a 
magnetic field, the $U/V (=2.39)$ will give rise to a 2D SDW/CDW-like 
modulations while a stronger $U/V (=2.44)$ would make the induced SDW/CDW 
to have one dimensional stripe-like structures. Then the spatial 
distributions of the LDOS as a function at various bias energies will be 
calculated and compared with those of the CDW. The spatial distributions 
of the integrated LDOS have also been obtained, and the result for 
$U/V=2.39$ shows strong LDOS intensity near the vortex core and its 
distribution seems to be round, not the four-fold symmetry as expected for 
the CDW and dSC orders. The result for $U/V=2.44$ shows that the 
stripe-like modulations are still existing but they are more localized 
near the vortex core and this is different from the CDW where the stripes 
are extended over the whole magnetic unit cell. These features are in good 
agreement with the experiments of Pan $et\;al.$ \cite{pan85} and Hoffman 
$et\;al.$ 
\cite{hoffman295}.

We start with an effective mean-field $t-t''-U-$V Hamiltonian in the mixed
state by assuming that the on-site repulsion $U$ is responsible for the
competing antiferromagnetism and the nearest-neighbor attraction $V$
causes the $d$-wave superconducting pairing.
\begin{eqnarray}
{\bf H}&=&-\sum_{{\bf ij}\sigma} t_{\bf ij} c_{{\bf
i}\sigma}^{\dagger}c_{{\bf j}\sigma}
+\sum_{{\bf i}\sigma} ( U\langle n_{{\bf i}\bar{\sigma}}\rangle - \mu )
c_{{\bf i}\sigma}^{\dagger}c_{{\bf i}\sigma} \nonumber \\
&&+\sum_{\bf ij} (\Delta_{\bf ij} c_{{\bf i}\uparrow}^{\dagger}
c_{{\bf j}\downarrow}^{\dagger} +\Delta_{\bf ij}^{*} c_{{\bf
j}\downarrow} c_{{\bf i}\uparrow} )\;,
\end{eqnarray}
where $t_{\bf ij}$ is the hopping integral, $\mu$ is the chemical
potential, and $\Delta_{\bf ij}=\frac{V}{2}\langle c_{{\bf i}\uparrow} 
c_{{\bf j}\downarrow}-c_{{\bf i}\downarrow}c_{{\bf j}\uparrow}\rangle$ is 
the spin-singlet $d$-wave bond order parameter.  The Hamiltonian above 
shall be diagonalized by using Bogoliubov-de Gennes' (BdG) equations,
\begin{eqnarray}
\sum_{\bf j}^N \left(\begin{array}{cc}
 {\cal H}_{{\bf i}j\sigma} & \Delta_{\bf ij} \\
 \Delta_{\bf ij}^* & -{\cal H}_{{\bf ij}\bar{\sigma}}^*
 \end{array}\right)
 \left(\begin{array}{c}
     u_{{\bf j}\sigma}^n \\
     v_{{\bf j}\bar{\sigma}}^n
 \end{array}\right)
 = E_n
 \left(\begin{array}{c}
     u_{{\bf i}\sigma}^n \\
     v_{{\bf i}\bar{\sigma}}^n
 \end{array}\right)\;,
\end{eqnarray}
where ${\cal H}_{{\bf ij}\sigma}=-t_{\bf ij} + ( U\langle n_{{\bf
i}\sigma}\rangle - \mu ) \delta_{\bf ij}$. Here, $t_{\bf ij} = \langle
t_{\bf ij} \rangle e^{i\varphi_{\bf ij}}\;.$
The Peierl's phase factor $\varphi_{\bf ij} =
\frac{\pi}{\Phi_0}\int_{r_{\bf i}}^{r_{\bf j}} {\bf A(r)}\cdot d {\bf
r}\;,$ with the superconducting flux quantum $\Phi_0=hc/2e$.
Within the Landau gauge ${\bf A(r)}=(-By,0,0)$, each magnetic unit cell 
can accommodate two superconducting vortices. The vortex carries a flux 
quantum $\Phi_0$ and locate at the center of a square area containing 
$N_x/2 \times N_y$ sites with $N_x=2\times N_y$.
Here, we choose the nearest-neighbor hopping $\langle t_{\bf ij}\rangle = 
t=1$ and the next-nearest-neighbor hopping $\langle t_{\bf ij}\rangle = 
t'=-0.25$ to match the curvature of the Fermi surface for most cuprate
superconductors. The exact diagonalization method to self-consistently
solve BdG equations with the periodic boundary conditions is employed to
get the $N$ positive eigenvalues $(E_n)$ with eigenvectors $(u_{{\bf
i}\uparrow}^n , v_{{\bf i}\downarrow}^n)$ and negative eigenvalues
$(\bar{E}_n)$ with eigenvectors $(-v_{{\bf i}\uparrow}^{n*} , u_{{\bf
i}\downarrow}^{n*} )$. The self-consistent conditions are
\begin{eqnarray}
\langle n_{{\bf i}\uparrow} \rangle &=&
  \sum_{n=1}^{2N}\left|{\bf u}_{\bf i}^n\right|^2 f(E_n)\;,\;
\langle n_{{\bf i}\downarrow} \rangle =
  \sum_{n=1}^{2N}\left|{\bf v}_{\bf i}^n\right|^2 [1-f(E_n)]\;,
     \nonumber \\
\Delta_{\bf ij} &=& \sum_{n=1}^{2N} \frac{V}{4} ({\bf u}_{\bf i}^n {\bf
v}_{\bf j}^{n*} + {\bf v}_{\bf i}^{n*} {\bf u}_{\bf j}^n) \tanh
(\frac{\beta E_n}{2})\;,
\end{eqnarray}
where ${\bf u}_{\bf i}^n = (-v_{{\bf i}\uparrow}^{n*}, u_{{\bf
i}\uparrow}^n )$ and ${\bf v}_{\bf i}^n = (u_{{\bf i}\downarrow}^{n*},
v_{{\bf i}\downarrow}^n )$ are the row vectors, and $f(E)=1\slash(e^{\beta
E}+1)$ is Fermi-Dirac distribution function. Since the calculation is
performed near the optimally doped regime, the filling factor,
$n_f=\sum_{{\bf i}\sigma} \langle c_{{\bf i}\sigma}^\dagger c_{{\bf
i}\sigma} \rangle /N_xN_y$, is fixed to be $0.85$, i.e., the hole doping
$\delta=0.15$. Each time when the on-site repulsion $U$ is varied, the
chemical potential $\mu$ needs to be adjusted .

\begin{figure}[t]
\centerline{\epsfxsize=8.0cm\epsfbox{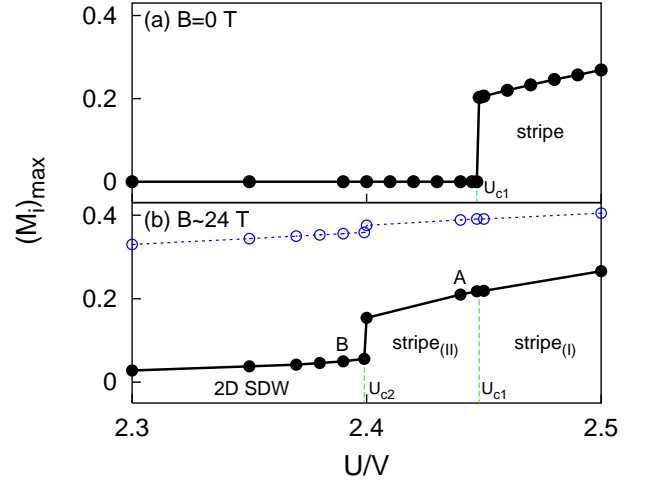}} \caption[*]{The maximum
value of the staggered magnetization in the zero-field (a), and in the
finite field (b). In (b), the open circle and the solid circle represent 
the (maximum) staggered magnetization at the center of the vortex core and 
away from the vortex cores, respectively. Stripe$_{(I)}$ is the region 
where the stripe modulations are intrinsic and stripe$_{(II)}$
corresponds to the region where the stripe modulations are field-induced.
The size of the unit cell is $N_x\times N_y = 48 \times 24$ correspoding 
to a magnetic field $B\sim24T$.}
\end{figure}

In the limit of $U/V\ll 1$, the system is in the state of pure dSC. In the
opposite limit $U/V\gg 1$, the system is the SDW states. However, the most
anomalous properties of cuprate superconductors do not correspond to these
extreme limits, but are in the intermediate case where both the SDW and the
SC may coexist. In order to simplify our discussion of the ratio of $U$ to 
$V$,
we set $V=1.0$. In Fig. 1(a), without the magnetic field, $B=0$, the
staggered magnetization ($M_{\bf i}$) or the SDW shows either the stripe
modulation or the uniform distribution with $M_{\bf i}=0$ in the background
of dSC depending upon the magnitude of $U$. For $U > U_{c1}$, the staggered
magnetization exhibits the stripe modulation with $8a$ as its periodicity
(a is the lattice constant). On the other hand, for the $U$ less than
$U_{c1}$, it shows the uniform distribution which is equivalent to the 
state of pure dSC. It is important to point here that a
two-dimensional (2D) SDW modulation can never be obtained when $B=0$ in 
the present self-consistent calculation. The transition between the 
SDW-stripe modulation and the uniform distribution is discontinuous. In 
Fig. 1(b), under an applied magnetic field, the staggered magnetization  
(solid line) displays the stripe modulation, the two dimensional SDW, or 
the uniform distribution depending on $U$. As $U$ is greater than 
$U_{c1}$, the stripe modulation existed in the zero field would be 
$slightly\;enhanced$ under a magnetic field (Stripe$_{(I)}$). In the 
region of $U_{c2} < U \leq U_{c1}$, the field induced staggered 
magnetization shows a stripe modulation which disappears in the zero field 
(Stripe$_{(II)}$). For $B \neq 0$, the transition between stripe$_{(I)}$ 
and stripe$_{(II)}$ is not apparent and only the slope shows a weak 
discontinuity. When $U \leq U_{c2}$, the field induced AF order changes 
from the stripe-like to a 2D SDW. If $U$ goes down far below $2.3$, the AF 
order could be completely suppressed both inside or outside the vortex 
cores. The transition between the stripe modulation and the 2D SDW is of 
the first order. Accompanying the stripe-like AF order, there also exists 
a CDW with the stripe modulation of the period $4a$. At the same time the 
dSC order parameter also acquires a CDW-like stripe modulation. In Fig. 
1(b), the staggered magnetization (open circles) at the vortex core center 
seems to be weakly $U$ dependent. In the following, $U=U_A=2.44$ and 
$U=U_B=2.39$ are chosen for our study of the LDOS. Both of these $U$ 
values would not generate the SDW/CDW order for nearly optimally doped
cuprate superconductors when $B=0$. In order to understand the 
characteristics between the two cases, we start with the LDOS formula
\begin{eqnarray}
\rho_{\bf i}(E) &=&-\frac{1}{M_x M_y} \sum_{n,{\bf k}}^{2N} \{
  \left| {\bf u}_{\bf i}^{n,{\bf k}} \right|^2 f'(E_{n,{\bf k}}-E)
 \nonumber \\
 &&+\left| {\bf v}_{\bf i}^{n,{\bf k}} \right|^2 f'(E_{n,{\bf k}} + E)
    \}\;,
\end{eqnarray}
where $\rho_{\bf i}(E)$ is proportional to the local differential tunneling
conductance as measured by STM experiment, and the summation is averaged
over $M_x \times M_y$ wavevectors in first Brillouin Zone.

The LDOS as a function of energy have been numerically calculated at the
vortex core center (solid line) and at site far away from the vortex
(dashed line) for $U_A$ and $U_B$, the results are respectively given in
Fig. 2(a) and 2(b). There the spatial profiles of the field induced CDW
modulations are presented in Fig. 2. For $U_A$, the field induced CDW has 
the stripe-like structure which extends over the whole magnetic unit cell 
with a period $4a$ while for $U_B$, the CDW modulation becomes two 
dimensional with four-fold symmetry.

Since the SDW gap develops inside the vortex core, the resonance peak
appeared in the LDOS \cite{wang52} near the zero bias in a pure dSC at the
core center is suppressed and splits into two peaks \cite{zhu87}. This
feature can be seen in the Fig. 2. Furthermore, STM experiments on YBCO
\cite{maggio75} and BSCCO \cite{pan85} measured the double peaks structure
within the maximum superconducting gap. The states associated with these
two peaks have been referred as the vortex core states. Here we would
like to point out that our results for the LDOS seem to agree better with
YBCO \cite{maggio75} than BSCCO \cite{pan85}. In Fig. 2, those vortex core
states with negative energies are centered at $E_{A}=-0.19$ with a width
$\Delta E_{A}=0.08$ and $E_{B}=-0.21$ with $\Delta E_{B}=0.04$ for $U_A$ 
and $U_B$, respectively.

\begin{figure}[t]
\centerline{\epsfxsize=8.0cm\epsfbox{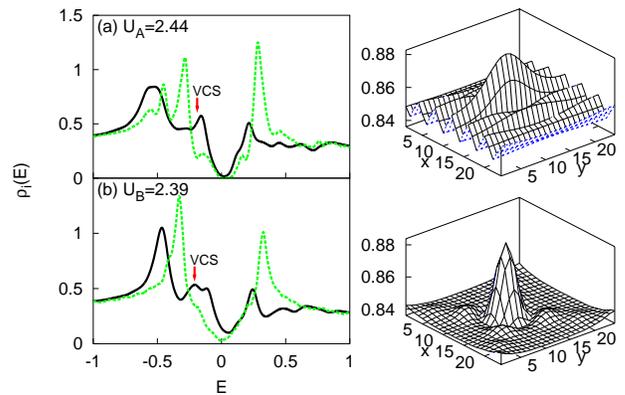}} \caption[*]{ The LDOS as a
function of energy, left panel, at the vortex core center for (a)
$U_A=2.44$, (b) $U_B=2.39$. The solid line is at the vortex core center,
and the dash line is at the site far away the vortex. The arrow points the
vortex core states (VCS). The spatial profiles of the corresponding CDW are
shown in the right panels of Fig. 2. The wavevectors in first Brillouin
Zone are $M_x\times M_y=24\times 24$. }
\end{figure}

Because the energy dependence of the LDOS at the vortex center does not
make any clear distinction between $U_A$ and $U_B$, we examine the
spatial profile of the LDOS at various energies and look for their
differences. In Fig. 3, the LDOS maps (the spatial distribution of the
LDOS), which are the LDOS with fixed energy at each site of the magnetic
unit cell, have been calculated at energies ranging from $0.0$ to $-0.4$
with $\delta E=0.01$ decrement. For $U_A$, as the energy far below the 
vortex core states (VCS) and close to the zero bias, the pattern in Fig. 
3(a) shows that the stripe structure with periodicity $4a$ is strongly 
localized near the vortex core, and its strength drops dramatically away 
from the vortex. When the energy near the VCS, such as Fig. 3(b), the 
stripe modulation in the LDOS extends over the whole unit cell, similar to 
the feature in CDW in Fig. 2(a). If the bias voltage goes above or 
becomes more negative than the VCS, such as Fig. 3(c), the stripe 
modulation is still extensive, but the intensities inside the vortex core 
are depressed than those outside the vortex core. Thus the STM spectra or 
the LDOS do not have all the features of the CDW.

\begin{figure}[t]
\centerline{\epsfxsize=8.0cm\epsfbox{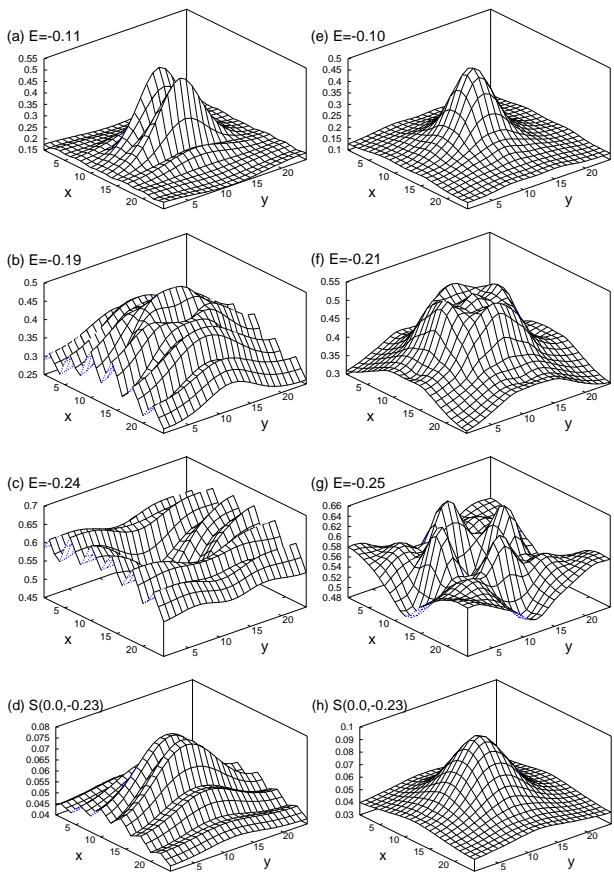}} \caption[*]{ The LDOS maps
are at energies (a) and (e)-close to the zero bias, (b) and (f)-near the
VCS, and (c) and (g) above the VCS. The integrated spectrum of LDOS
$S(E_1,E_2)$ of (d) and (h) are from the Fermi energy to the upper bound
of the VCS, i.e., $S(0.0,-0.23)$. The left and right panel are for
$U_A=2.44$ and $U_B=2.39$, respectively. The wavevectors in first
Brillouin Zone are $M_x\times M_y=6\times 6$.}
\end{figure}

For $U_B$, as the energy far below the vortex core states (VCS) [see Fig.
3(e)], the LDOS pattern shows a $round$ bump with the size of a vortex
core. When the energy is near the VCS, such as in Fig. 3(f), the LDOS shows
a distribution with rather weak oscillations, and its feature is not 
quite similar to that of the CDW in Fig. 2(b). If the energy is above the 
VCS, such as that in Fig. 3(g), the modulation here clearly displays the 
pattern with $4$-fold symmetry. However, the STM images obtained 
experimentally \cite{hoffman295} are results of integrating the spectral 
density between energies $E_1$ and $E_2$, which is defined as
\begin{eqnarray}
S(E_1,E_2)=\sum_{E_1}^{E_2} \rho_{\bf i}(E) \delta E\;,
\end{eqnarray}
There, \cite{hoffman295} $E_1$ is taken to be $0$ and $E_2$ has been set
near the energy of the VCS below the chemical potential. In Fig. 3(d), the
integrated spectrum $S(0.0,-0.23)$ of the LDOS is obtained by summing the
LODS from $E_1$ to $E_2$ with an energy spacing $0.01$. It shows that the
intensity of the stripe modulations is dominantly concentrated near the
vortex core and decays rapidly when away from the vortex. This feature
originates from the behaviors of the LDOS at energies with magnitudes
smaller than those of the VCS [see Fig. 3(a)]. Since the LDOS maps display
the properties of the $eigenfunction$ at the energy $E$, the checkerboard
pattern observed by the experiments could be explained in terms of the
superposition of the degenerate eigenfunctions describing the stripe
modulations along $x$- and $y$- directions. All these behaviors are in 
good agreement with the experiment \cite{hoffman295}. We expected that the 
qualitative features obtained for $U_A$ should still remain even if $U > 
U_{c1}$ and the stripe phase is intrinsic not magnetic field induced (see 
Fig. 1). In Fig. 3(h), the integrated spectrum of the LDOS for $U_B$ 
exhibits a $round$ bump over the vortex core region which is different 
from that of the CDW as shown in Fig. 2(b). For a sample without the 
stripe modulations, the profile of the integrated LDOS near the vortex 
core seems to be rather $round$ and does not possess the strong $4$-fold 
symmetry as expected for a dSC \cite{zhu87}. This feature is also 
consistent with the STM experimental measurements \cite{pan85} provided 
that the samples used there is different from that of \cite{hoffman295}.

In conclusion, we have numerically investigated the interplay between the
dSC and the competing SDW/CDW orders by varying the strength of $U/V$ for
samples close to the optimal doping. For finite field, we show that both
stripe-like and two dimensional SDW/CDW orders, depending on the magnitude
of $U/V$, may be induced in the background of the dSC. We in particular
calculate the spatial distribution of the LDOS with and without the SDW/CDW
stripes at various bias energies. We point out that the checkerboard
pattern near the vortex core observed by Hoffman $et\;al.$
\cite{hoffman295} could be interpreted in terms of superposition of the
field-induced $x$- and $y$- oriented stripes. The obtained features of our
integrated LDOS with stripes and without stripes [see Figs. 3(d) and 3(h)]
are in good agreement with the STM experiments \cite{hoffman295,pan85}. It
is well known that the theoretically obtained profiles for the dSC order
parameters together with the induced 2D SDW/CDW modulations near the vortex
core exhibit a clear 4-fold symmetry. This symmetry, however, so far has
not been confirmed by existing STM measurements. Here, we predict that 
this 4-fold symmetry should be more easily detectable when the bias is 
placed beyond the energies of the vortex core states [see Fig. 3(f)].

${\bf Acknowledgements}$: We wish thank S.H. Pan for  useful comments and
suggestions. This work is supported by the Texas Center for Superconductivity
and Advanced Material and Department of Physics at the University of Houston,
and by the grant of the Robert A. Welch Foundation.

\end{document}